\documentclass[twocolumn]{autart}                                     % preceding line to obtain a two-column
\usepackage{ntheorem}
\usepackage[numbers,sort&compress]{natbib}
\usepackage{amsfonts}
\usepackage{mathrsfs}
\usepackage{amsmath,amssymb}
\usepackage{graphicx}% for pdf, bitmapped graphics files
\usepackage{amsmath} % assumes amsmath package installed
\usepackage{amssymb}  % assumes amsmath package installed
\usepackage{color}
\usepackage{booktabs}
\usepackage{colortbl}           % Include this line if your
\renewcommand{\theequation}{\arabic{section}.\arabic{equation}}

\usepackage{subfigure}
\usepackage{setspace}
\allowdisplaybreaks
\newtheorem{defx}{Definition}
\newtheorem{assumption}{Assumption}
\newtheorem{lemmax}{Lemma}
\newtheorem{corollaryx}{Corollary}
\newtheorem{remarkx}{Remark}
\newtheorem{example}{Example}
\newtheorem{propertyx}{Property}
\newtheorem{theoremx}{Theorem}
\newtheorem*{proof}{Proof}
\begin{document}

\begin{frontmatter}

\title{Predefined-time Terminal Sliding Mode Control of\\ Robot Manipulators\thanksref{footnoteinfo}} % Title, preferably not more
                                                % than 10 words.

\thanks[footnoteinfo]{This work was supported by the National Natural Science Foundation of China under Grant 61703374,
and the Fundamental Research Founds for National University, China University of Geosciences (Wuhan) under Grant 1910491B05.\\
$^*$ Corresponding author: Ming-Feng Ge.}

\author[CUG]{Chang-Duo Liang}\ead{liangchangduo93@163.com},
\author[CUG]{Ming-Feng Ge$^*$}\ead{fmgabc@163.com},
\author[Hust,KL]{Zhi-Wei Liu}\ead{zwliu@hust.edu.cn},
\author[Hust,KL]{Yan-Wu Wang}\ead{wangyw@hust.edu.cn},              % e-mail address
\author[MLLG]{Hamid Reza Karimi}\ead{hamidreza.karimi@polimi.it}  % (ead) as shown

\address[CUG]{School of Mechanical Engineering and Electronic Information,
              China University of Geosciences, Wuhan 430074, China }
\address[Hust]{School of Artificial Intelligence and Automation,
Huazhong University of Science and Technology, Wuhan 430074, China}
\address[KL]{Key Laboratory of Image Processing and Intelligent Control,
Ministry of Education, Huazhong University of Science and Technology,
Wuhan 430074, China}
\address[MLLG]{Department of Mechanical Engineering, Politecnico di Milano, 20156 Milan, Italy}        % here.

                             % keyword list or with the
                                          % help of the Automatica
                                          % keyword wizard

\begin{abstract}                          % Abstract of not more than 200 words.
In this paper,
we present a new terminal sliding mode control to achieve predefined-time stability of robot manipulators.
The proposed control is developed based on a novel predefined-time terminal sliding mode (PTSM) surface,
on which the states are forced to reach the origin in a predefined time,
i.e., the settling time is independent to the initial condition and can be explicitly user-defined via adjusting some specific parameters called the \textit{predefined-time parameters}.
It is also demonstrated that the proposed control can provide satisfactory steady-state performance in the case of both external disturbances and parametric uncertainties.
Besides, we present a formal systemic analysis method to derive the sufficient conditions for guaranteeing the predefined-time convergence of the closed-loop system.
Finally, the effectiveness and performance of the presented control scheme are illustrated through both theoretical comparisons and numerical simulations.
\end{abstract}

\end{frontmatter}

\section{Introduction}
In the past decades,
the stabilization problems of nonlinear systems have been a magnet for large research attention due to its wide potential applications \cite{ShiPeng01, LiFB01, ShiPeng02}.
These results were mainly focused on first-, second-order nonlinear systems and Lipschitz-type systems.
Thus, they cannot be directly applied to regulate the robot manipulators, whose dynamics are generally described as the Euler-Lagrange system.
This system has been recognized as a practical model to describe various rigid bodies,
such as surface vehicles \cite{Karimi01},
networked robotic systems \cite{Ge02,LiangCD},
and teleoperation systems \cite{WangZ2019}.

Therefore, increasing efforts have been devoted to the stabilization of robot manipulators and other Euler-Lagrange systems \cite{MengZY2013, Ge03, SunYC2017, HeWei2015}.
In these systems, the parametric uncertainties and external disturbances are inevitably involved within the dynamics and generally have negative impacts on system stabilization.
To eliminate these impacts, there have arisen many control techniques,
including the parameter-adaptive sliding mode control \cite{LiangCD, MengZY2013},
neural-network-based sliding mode control \cite{SunYC2017, HeWei2015}, to name a few.
It has been illustrated from the above-mentioned literatures that the sliding mode control is an effective way to deal with such class of robust stabilization problems due to its capability of disturbance rejection.

Most of the existing sliding mode control schemes are designed based on the linear sliding mode (LSM) surface,
on which the states converge to the origin exponentially as time approaches infinity \cite{Ge02,LiangCD}.
Then, in order to achieve \textit{finite-time stability} of the states,
the terminal sliding mode control technique has been presented based on terminal sliding mode (TSM) surface,
which can force the states to approach the origin in a finite time \cite{MengZY2010, FengYong01, Yu01}.
The obtained finite time is related to the initial condition, detailedly, it generally increases if the value of the initial states rises, which thus prevents the applications of the finite-time TSM control in the case of large initial values.
%The preceding sliding mode controller for Euler-Lagrange systems are mostly designed by employing the linear sliding surface \cite{LiangCD, MengZY2013, WangHanLei2015},
%which eventually derives the asymptotic convergence of the states.
%Nevertheless, in practical applications,
%convergence rate is regarded as a important index to evaluate the working performance and effectiveness of the proposed control algorithm.
%It then motivates groups of researches on finite-time tracking control \cite{MengZY2010, Ge03},
%which generates much faster convergence rate and higher tracking precision.
%As such, Feng, et al. \cite{FengYong01} designed a kind of non-singular terminal sliding mode (TSM) surface for second-order uncertain nonlinear system,
%based on which the non-singular TSM controller was further proposed to solve the finite-time tracking problem of rigid manipulators.
%Later, a continuous sign-function-based TSM surface was designed in \cite{Yu01}, and the continuous finite-time control for manipulators had been investigated.
%It is noteworthy that the settling time function obtained in the above literatures rely on the inial values of the states,
%which prohibits their practical applications in the case that the inial values are unknown.
Thus, more recent researches focused on deriving the uniform boundedness of
the settling time regardless of the initial conditions, being referred to as \textit{fixed-time stability} \cite{ZuoZY2015, ZuoZY2014}.
Although the fixed-time schemes can generally provide faster convergence speed comparing with the finite-time ones,
the obtained fixed time generally has a complicated relationship with several system/control parameters, and cannot be easily user-defined.
However, it is of great significance to define the settling time in advance for task planning of practical engineering applications.
Therefore, to predefine the settling time,
a novel concept named \textit{predefined-time stability} has been introduced to describe a new type of stability,
in which the settling time for the convergence of the states can be explicitly user-defined
through the adjustment of some \textit{predefined-time parameters} \cite{Munoz-Vazquez, Munoz-Vazquez02, Sanchez-Torres, Aldana-Lopez}.

As a consequence, the prescribed-time (i.e., predefined-time) stability of first-order integrator has been presented in \cite{WangY2018} by introducing a time-varying scaling function.
Wang, et al, \cite{WangZ2019} have investigated the adaptive fault-tolerant prescribed-time control for teleoperation systems by employing a specific time-varying piece-wise function.
Due to the employment of the LSM surface, the practical predefined-time tracking problems (i.e., the tracking errors converge to a bounded set) of robotic manipulator has been solved in \cite{Munoz-Vazquez}.
All in all, the zero-error predefined-time stabilization problem for robot manipulators has not been well addressed,
due to the lack of an applicable predefined-time terminal sliding mode (PTSM) surface,
on which the states will be forced towards the origin within a predefined time.

Motivated by the above discussions,
a newly-designed PTSM surface is proposed to handle the predefined-time stabilization problem of robot manipulators with both external disturbances and parametric uncertainties.
The main contribution of this paper is to solve the zero-error predefined-time stabilization problem of robot manipulators, different from the results presented in \cite{Munoz-Vazquez}, which can only achieve practical predefined-time stability of the states, i.e., the states converge to a bounded neighborhood of the origin in a predefined time.
We present a new PTSM to replace the LSM used in \cite{Munoz-Vazquez} to solve such a challenging problem.
It is also illustrated from both numerical simulations and theoretical comparisons that the presented PTSM control schemes can provide satisfactory performance (i.e., the cost of the control input, the steady-state performance) and can be easily extended to develop controllers for effectively regulating other higher-order nonlinear systems in a predefined time.

The remaining parts are organized as follows.
Section \ref{sec2} provides the relative preliminaries.
The newly-designed PTSM surface and the predefined-time stability are presented in Section \ref{sec3}.
The predefined-time stabilization problem of robot manipulators is analyzed in Section \ref{sec4}.
The numerical simulation results are included in Section \ref{sec5}.
Finally, conclusions are summarized in Section \ref{sec6}.

\textit{Notations:}
$\mathbb R$ denotes the real number field, and $\mathbb R^n$ represents the $n$-dimensional Euclidean space.
$\lambda_{\min}(\cdot)$ symbolizes the minimum eigenvalue of the corresponding matrix.
$\circ$ denotes the Hadamard product, detailedly, given $x = [x_1, x_2, \ldots, x_n]^T$ and $y = [y_1, y_2, \ldots, y_n]^T$, one has $x \circ y = [x_1y_1, x_2y_2, \ldots, x_ny_n]^T$.
Besides, some operate modes are defined as follows,
${\rm sig} (x)^k = [{\rm sgn}(x_1)|x_1|^k, {\rm sgn}(x_2)|x_2|^k, \ldots, {\rm sgn}(x_n)|x_n|^k]^T$,
and ${\left[\kern-0.15em\left[ x \right]\kern-0.15em\right]^k} = [x_1^k, x_2^k, \ldots, x_n^k]^T$,
where $k$ is a positive constant.

\section{Preliminaries }\label{sec2}

\subsection{Conventional terminal sliding mode surface}
The conventional TSM surfaces for deriving \textit{finite-time stability} \cite{Yu01} and \textit{fixed-time stability} \cite{ZuoZY2015, ZuoZY2014} are usually respectively designed as
\begin{align}
&s = \dot x + b_1{\rm sig}(x)^\nu\\
&s = \dot x + a_1x + b_1 {\rm sig}(x)^\nu,
\end{align}
and
\begin{align}
s = \dot x + a_2 x^{\frac{m_1}{n_1}} + b_2 x^{\frac{m_2}{n_2}},
\end{align}
where $x \in \mathbb R$, $a_1, b_1 > 0$, $0< \nu <1$, $a_2, b_2>0$, $m_1, n_1, m_2$ and $n_2$ are positive odd integers with $m_1 > n_1$ and $m_2 < n_2 $.

\vspace{0.5em}
\begin{remarkx}
According to the definitions of finite-time stability \cite{HongH2001} and fixed-time stability \citep[Lemma 2]{ZuoZY2014},
it can be derived that $x$ converges to the origin in finite-time and fixed-time after reaching the sliding surface (i.e., $s = 0$),
and the settling time functions are respectively given as follows:
\begin{align}
&\mathcal T \leq \frac{1}{b_1(1-\nu)}|x(t_0)|^{1-\nu},\\
&\mathcal T \leq \frac{1}{a_1(1-\nu)} \ln \frac{a_1|x(t_0)|^{1-\nu} + b_1}{b_1},\\
&\mathcal T \leq \frac{n_1}{a_2(m_1-n_1)} + \frac{n_2}{b_2(n_2-m_2)}.
\end{align}
\end{remarkx}

\vspace{0.5em}
\begin{remarkx}
Using the conventional TSM,
it is noteworthy that the settling time functions are extremely complicated.
Then, it becomes difficult to derive the desired uniform settling time by adjusting the control parameters,
which thus motivates us to explore a novel PTSM surface to solve the above problem.
\end{remarkx}

\subsection{Lemmas}
Some useful lemmas are given as follows.

\vspace{0.5em}
\begin{lemmax}\label{L1}\cite{Munoz-Vazquez}
For given system $\dot x = f(t,x)$,
if there exists a positive-definite Lyapunov function $V(x)$ such that
\begin{align}
\dot V(x) \le  - \frac{\pi }{{\rho {{\mathcal T}_c}}}\left( {{V^{1 - \frac{\rho }{2}}}(x) + {V^{1 + \frac{\rho }{2}}}(x)} \right),
\end{align}
where $\mathcal T_c >0$ and $0< \rho <1$,
then the origin is a globally predefined-time stable equilibrium of the considered system with $\mathcal T_c$ being the predefined time,
namely, $x$ and $\dot x$ converges to zero within $t \leq \mathcal T_c$.
\end{lemmax}

\vspace{0.5em}
\begin{lemmax}\label{L2} \cite{NingBD}
Consider the differential system $\dot x = -\phi(t) x$, $x(0) = x_0$,
and chose
\begin{equation}
\phi(t) = \frac {\dot \varepsilon(t)} {1 - \varepsilon(t) + \epsilon},
\end{equation}
where $0 < \epsilon \ll 1$, and $\varepsilon(t)$ is the time base generator (TBG) which satisfying the following properties:
\begin{enumerate}
\item[(\romannumeral1)] $\varepsilon(t) \in C^p$ on $(0, +\infty)$, where $p \in \mathbb N$ and $p \geq 2$, i.e.,  $\varepsilon(t)$ is continuous and at least second-order derivable on $(0, +\infty)$;
\item[(\romannumeral2)] $\varepsilon(0) = 0$, $\varepsilon(\mathcal T_c) = 1$, $\dot \varepsilon(0)= \dot \varepsilon(\mathcal T_c) = 0$, and $\varepsilon(t) = 1$, $\dot \varepsilon(t) = 0$ when $t > \mathcal T_c$, where $0 < \mathcal T_c < +\infty$ is a predefined time instant.
\item[(\romannumeral3)] $\varepsilon(t)$ is non-decreasing on $[0,\mathcal T_c]$.
\end{enumerate}
Then the state $x$ reaches $\frac {\epsilon} {1+\epsilon} x_0$ at $\mathcal T_c$ which is regardless of the initial condition.
\end{lemmax}

\section{The predefined-time terminal sliding mode surface and the predefined-time stability}\label{sec3}

\subsection{The design of the PTSM surface}
In this subsection, we are going to propose a PTSM surface,
under which the upper bound of the settling time appears explicitly as a predefined-time parameter in the control design.
The PTSM with predefined-time convergence is described as
\begin{equation}\label{e1}
s = \dot x + \frac{{{(1 + {x^2})}^{\frac{3}{2}}} }{{\mathcal T_s (1-\gamma) }}{\rm sig}{\left( {\frac{x}{{\sqrt {1 + {x^2}} }}} \right)^\gamma },
\end{equation}
where $x,\dot x \in \mathbb R$, $\mathcal T_s > 0$, and $0 < \gamma < 1$.

\vspace{0.5em}
\begin{theoremx}\label{T1}
Designing PTSM surface as (\ref{e1}), the achievement of $s = 0$ guarantees that the predefined-time convergence to the origin of the states with $\mathcal T_s$ being the predefined-time parameter,
namely, once the PTSM surface is reached, $x$ and $\dot x$ converge to zero within $t \leq \mathcal T_s$.
\end{theoremx}

\vspace{0.5em}
\begin{proof}
Once $s = 0$ is derived, one has
\begin{equation}
\dot x =  - \frac{{{(1 + {x^2})}^{\frac{3}{2}}} }{{{{\mathcal T}_s} (1-\gamma) }} {\rm sig} {\left( {\frac{x}{{\sqrt {1 + {x^2}} }}} \right)^\gamma }.
\end{equation}
\textbf{In the case that} $\boldsymbol{x \geq 0}$, the above differential system turns to be
\begin{equation}
\dot x =  - \frac{{{(1 + {x^2})}^{\frac{3}{2}}} }{{{{\mathcal T}_s}(1 - \gamma )}} {\left( {\frac{x}{{\sqrt {1 + {x^2}} }}} \right)^\gamma }.
\end{equation}
Construct the Lyapunov function candidate as $V(x) = {\left( {x}/{\sqrt {1 + {x^2}} } \right)^{1-\gamma} }$,
which is obviously positive-definite.
It then follows that
\begin{align}
\dot V(x) &= \frac{{(1 - \gamma )\dot x}}{{{(1 + {x^2})}^{\frac{3}{2}}} }{\left( {\frac{x}{{\sqrt {1 + {x^2}} }}} \right)^{ - \gamma }} = - \frac{1}{{{{\mathcal T}_s}}}.
\end{align}
Then, based on \textit{Lyapunov stability},
the states converge to zero in finite time, and the settling time function is given as $\mathcal T \leq V(x(t_0)){\mathcal T_s}$.
Note that $0 \leq x/\sqrt{1 + x^2} < 1$ on $x \in [0,+\infty)$.
It follows that $0 \leq V(x(t_0)) <1$,
and thus one has $\mathcal T \leq \mathcal T_s$ which does not depend on the initial value of $x$.
The same conclusion can be derived by similar analysis \textbf{in the case that} $\boldsymbol{x < 0}$.
This completes the proof.
\end{proof}

\subsection{Predefined-time stability for second-order systems}\label{sec2.2}
\begin{figure}[t]
\centering
\makebox{\includegraphics[width=6cm]{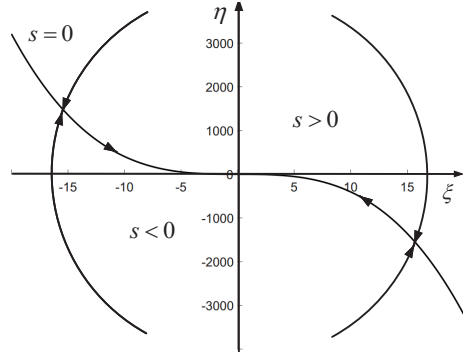}}
\caption{\label{F-8} The phase plot of the system.}
\end{figure}

In this subsection,
the predefined-time stabilization problem for the uncertain second-order system is first analyzed.
Consider the uncertain second-order system described as
\begin{equation}\label{e2}
\left\{ \begin{array}{l}
\dot \xi (t) = \eta (t),\\
\dot \eta  = \tau (t) + f(t,\xi (t),\eta (t)),
\end{array} \right.
\end{equation}
where $\xi, \eta \in \mathbb R^n$ are the state vectors,
$\tau(t)$ is the control input,
and $f(t,\xi (t),\eta (t))$ denotes the uncertain term including the parametric uncertainties and external disturbances.
A reasonable assumption on $f(t,\xi (t),\eta (t))$ is presented as follows.

\vspace{0.5em}
\begin{assumption}
The uncertain term $f(t,\xi (t),\eta (t))$ are bounded,
i.e., there exists a positive scalar $\overline \sigma_f$ such that $\|f(t,\xi (t),\eta (t))\| \leq \overline \sigma_f$.
\end{assumption}

\vspace{0.5em}
\begin{defx}
The main problem considered here is to design proper control input $\tau(t)$ such that system (\ref{e2}) is \textit{predefined-time stable},
namely, there exists an \textit{predefined-time parameter} $\mathcal T_f$ in the control design such that
\begin{equation}
\left\{ \begin{array}{l}
\mathop {\lim }\limits_{t \to {{\mathcal T}_f}} \xi (t) = 0,\\
\mathop {\lim }\limits_{t \to {{\mathcal T}_f}} \eta (t) = 0,
\end{array} \right.
\end{equation}
and $\xi (t) = 0$, $\eta (t) = 0$ when $t \geq \mathcal T_f$.
\end{defx}

Firstly, the PTSM surface is designed as
\begin{equation}
s = \eta  +\! \frac{{{{\left[\kern-0.30em\left[ {{\textbf{1}_n} \!+\! {{\left[\kern-0.15em\left[ \xi
 \right]\kern-0.15em\right]}^2}}
 \right]\kern-0.30em\right]}^{\frac{3}{2}}}}}{{{{\mathcal T}_s}(1 - \gamma )}} \!\circ {\rm sig}{\left( {\xi \circ\! {{\left[\kern-0.30em\left[ {{\textbf{1}_n} \!+\! {{\left[\kern-0.15em\left[ \xi
 \right]\kern-0.15em\right]}^2}}
 \right]\kern-0.30em\right]}^{ - \frac{1}{2}}}} \right)^\gamma },
\end{equation}
where $\mathcal T_s > 0 $ and $0 < \gamma <1$.
The control input for (\ref{e2}) is designed as $\tau = \tau_{eq} + \tau_s$,
where $\tau_{eq}$, $\tau_s$ respectively represent the equivalent control law and the nonlinear hitting control law, and are concretely presented as follows:
\begin{align}
&{\tau _{eq}} =  - \frac{{3\xi  \!\circ\! {{\left[\kern-0.30em\left[ {{\textbf{1}_n} \!+\! {{\left[\kern-0.15em\left[ \xi
 \right]\kern-0.15em\right]}^2}}
 \right]\kern-0.30em\right]}^{\frac{1}{2}}} \!\!\circ \eta }}{{{{\mathcal T}_s}(1 - \gamma )}} \circ {\rm sig}{\left( {\xi \!\circ\! {{\left[\kern-0.30em\left[ {{\textbf{1}_n} \!+\! {{\left[\kern-0.15em\left[ \xi
 \right]\kern-0.15em\right]}^2}}
 \right]\kern-0.30em\right]}^{ - \frac{1}{2}}}} \right)^\gamma } \nonumber\\
&\;\qquad- \frac{{\gamma \eta }}{{{{\mathcal T}_s}(1 - \gamma )}} \circ {\left[\kern-0.35em\left[ {\left| \xi \right| \circ\! {{\left[\kern-0.30em\left[ {{\textbf{1}_n} \!+\! {{\left[\kern-0.15em\left[ \xi
 \right]\kern-0.15em\right]}^2}}
 \right]\kern-0.30em\right]}^{ - \frac{1}{2}}}} \right]\kern-0.35em\right]^{\gamma  - 1}}, \label{e4}\\
&{\tau _s} =  - \frac{\pi }{{\rho {{\mathcal T}_c}}}\left( {1 + {{\left\| s \right\|}^{2\rho} }} \right)\frac {s}{{{\left\| s \right\|}^\rho }} - {\rm K}_f {\rm sgn} (s)\label{e5},
\end{align}
where $\mathcal T_c >0$, $0< \rho < 1$ and ${\rm K}_f$ is a diagonal positive-definite gain matrix.

Substituting the designed control input $\tau$ into (\ref{e2}) and differentiating $s$ along (\ref{e2}) yield the following closed-loop system,
\begin{equation}\label{e3}
\dot s =  - \frac{\pi }{{\rho {{\mathcal T}_c}}}\left( {1 + {{\left\| s \right\|}^{2\rho} }} \right)\frac {s}{{{\left\| s \right\|}^\rho }} - {\rm K}_f {\rm sgn} (s) + f(t,\xi (t),\eta (t)).
\end{equation}
Further, we focus on the stability analysis of (\ref{e3}),
and the corresponding theorem is presented as follows.

\vspace{0.5em}
\begin{theoremx}\label{T2}
Utilizing the designed control input (\ref{e4}) and (\ref{e5}) for the considered uncertain system (\ref{e2}),
if
\begin{align}
&\mathcal T_c >0, \quad \mathcal T_s >0 ,\\
&\lambda_{\min}({\rm K}_f) \geq \overline \sigma_f,\label{e6}
\end{align}
then (\ref{e2}) is predefined-time stable with the predefined-time parameter being $\mathcal T_f = \mathcal T_c +\mathcal T_s$,
namely $\xi$, $\eta$ converge to the origin within $t \leq \mathcal T_f$.
\end{theoremx}

\vspace{0.5em}
\begin{proof}
Construct the Lyapunov function candidate for (\ref{e3}) as $V(s) = \frac {1}{2}s^T s$.
It then follows that
\begin{align}
V(s) =& -\! \frac{\pi }{{\rho {{\mathcal T}_c}}}\!\!\left(\! {{{\left\| s \right\|}^{2 \!-\! \rho }} \!\!+\! {{\left\| s \right\|}^{2 \!+\! \rho }}} \!\right) \!-\! {s^T}{{\rm K}_f}{\rm{sgn}}(s) \!+\! {s^T}f \nonumber \\
  \leq& -\! \frac{\pi }{{\rho {{\mathcal T}_c}}}\!\!\left(\! {{{\left\| s \right\|}^{2 \!-\! \rho }} \!\!+\! {{\left\| s \right\|}^{2 \!+\! \rho }}} \!\right) \!\!-\! \left( {{\lambda _{\min }}({\rm K}_f) \!-\!  \left\| {f} \right\|} \right)\!\left\| s \right\| \nonumber \\
     \leq&  - \frac{\pi }{{\rho {{\mathcal T}_c}}}\!\!\left( {{V^{1 - \frac{\rho }{2}}}(s) + {V^{1 + \frac{\rho }{2}}}(s)} \right),
\end{align}
where $\rho \in (0,1)$ is predefined right after (\ref{e5}),
(\ref{e6}) has been utilized to derive the above inequality.
It thus follows from Lemma \ref{L1} that $s$ converge to the origin within predefined time $\mathcal T_c$.
Then, based on the the employment of the PTSM surface and the results in Theorem \ref{T1},
one can eventually obtain that $\xi$ and $\eta$ converge to zero within $t \leq \mathcal T_f = \mathcal T_c + \mathcal T_s$.
This completes the proof
\end{proof}

\vspace{0.5em}
\begin{remarkx}
The phase plot of system (\ref{e2}) using the designed algorithm is shown in Fig.\ref{F-8}.
For any chosen initial values,
the trajectory $\xi(t)$ will touch the PTSM surface $s = 0$,
and then converge to the origin along the PTSM surface monotonically within a predefined time.
\end{remarkx}

\section{Predefined-time tracking control for robot manipulators}\label{sec4}
In this section, based on the aforementioned predefined-time stability of the uncertain second-order system,
we will further study the predefined-time stabilization problem of robot manipulators.
To be specific, the \textit{main objective} is to design proper PTSM control schemes to actuate the robot manipulator to track the desired trajectory.

\subsection{System and problem formulation}\label{sec3.1}
The dynamics of the robot manipulator with $n$ degrees of freedoms (DOFs) can be described by the following Euler-Lagrange system,
\begin{equation}\label{e7}
M(q){\ddot q} + C(q, \dot q){\dot q} + g(q) = \tau(t) + d(t),
\end{equation}
where $q$, $\dot q \in \mathbb R^n$ stands for the generalized-joint coordinates and velocities,
$M(q) = M_0(q) + \Delta M(q) \in \mathbb R^{n \times n}$ denotes the inertia matrix,
$C(q, \dot q) = C_0(q, \dot q) + \Delta C(q, \dot q) \in \mathbb R^{n \times n}$ represents the Coriolis-centrifugal matrix,
$g(q) = g_0(q) + \Delta g(q) \in \mathbb R^n$ symbolizes the gravitational torque,
$d(t)$ denotes the external disturbance,
$\tau (t)$ is the torque input to be designed.
Besides, $M_0(q)$, $C_0(q, \dot q)$ and $g_0(q)$ are the nominal terms with respect to  $M(q)$, $C(q, \dot q)$ and $g(q)$ which can be used in the control design.
$\Delta M(q)$, $\Delta C(q, \dot q)$ and $\Delta g(q)$ denote the dynamical uncertainties.
Then (\ref{e7}) can be rewritten as the following form:
\begin{equation}\label{e10}
M_0(q){\ddot q} + C_0(q, \dot q){\dot q} + g_0(q) = \tau(t) + d(t) + h(t),
\end{equation}
where $h(t) = -\Delta M(q){\ddot q} - \Delta C(q, \dot q){\dot q} - \Delta g(q)$ denotes the uncertain term aroused by the dynamical uncertainties.
Besides, a reasonable assumption is given as follows.

\vspace{0.5em}
\begin{assumption}\label{A2}
The external disturbance $d(t)$ imposed on system (\ref{e7}) is upper-bounded,
namely, there exists positive $\overline \sigma_d$ such that $\|d(t)\| \leq \overline \sigma_d$.
\end{assumption}

Some properties of (\ref{e7}) are presented as follows.

\vspace{0.5em}
\begin{propertyx}\label{P1}
$M(q)$ and $M_0(q)$ are symmetric and positive definite.
Besides, $\dot M(q) - 2C(q, \dot q)$ and $\dot M_0(q) - 2C_0(q, \dot q)$ are skew symmetric,
namely, for any given $\delta  \in \mathbb R^n$, it can be derived that $\delta^T [ \dot M(q) - 2C(q, \dot q) ] \delta = 0$ and $\delta^T [ \dot M_0(q) - 2C_0(q, \dot q) ] \delta = 0$.
\end{propertyx}

\vspace{0.5em}
\begin{propertyx}\label{P2}
$M(q)$, $C(q, \dot q)$, and $g(q)$ are bounded for all possible $q$,
i.e., $\|M(q)\| \leq \overline \sigma_m$,
$\|C(q, \dot q)\| \leq \overline \sigma_c \|\dot q\|$,
and $\|g(q)\| \leq \overline \sigma_{g1} + \overline \sigma_{g2} \|q\|$,
where $\overline \sigma_m$, $\overline \sigma_c$, $\overline \sigma_{g1}$, and $\overline \sigma_{g2}$ are positive constants.
\end{propertyx}

On the other hand,
the desired trajectory is denoted by ${\rm col}(q_r, \omega_r, \alpha_r)$,
where $q_r$, $\omega_r$, and $\alpha_r \in \mathbb R^n$ denote the desired generalized-joint coordinate, velocity, and acceleration of the referenced trajectory,
i.e., ${{\dot q }_r} = {\omega_r}$, ${{\dot \omega }_r} = {\alpha_r}$.
A mild assumption on the acceleration of the referenced trajectory is presented as follows.

\vspace{0.5em}
\begin{assumption}\label{A3}
The acceleration of the referenced trajectory is bounded,
namely $\|\alpha_r\| \leq \overline \sigma_\alpha$,
where $\overline \sigma_\alpha$ is a positive constant.
\end{assumption}

\vspace{0.5em}
\begin{defx}
The \textit{zero-error predefined-time stabilization problem} for robot manipulators is solved if there exists an \textit{predefined-time parameter} $\mathcal T_f$ in the control design such that
\begin{equation}
\left\{ \begin{array}{l}
\mathop {\lim }\limits_{t \to {{\mathcal T}_f}} \|e\| = 0,\\
\mathop {\lim }\limits_{t \to {{\mathcal T}_f}} \|\dot e\| = 0,
\end{array} \right.
\end{equation}
and $\|e\| = 0$, $\|\dot e\| = 0$ for $t \geq \mathcal T_f$,
where $e = q - q_d$ and $\dot e = \dot q - \omega_d$ denote the errors states.
Especially, the \textit{practical predefined-time stabilization problem} is solved if
\begin{equation}
\left\{ \begin{array}{l}
\mathop {\lim }\limits_{t \to {{\mathcal T}_f}} \|e\| \leq \delta_1,\\
\mathop {\lim }\limits_{t \to {{\mathcal T}_f}} \|\dot e\| \leq \delta_2,
\end{array} \right.
\end{equation}
and $\|e\| \leq \delta_1$, $\|\dot e\| \leq \delta_2 $ for $t \geq \mathcal T_f$,
where $\delta_1$, $\delta_2 > 0$ can be adjusted by appropriate control parameters.

\end{defx}

\subsection{Control design and stability analysis }
Based on Property \ref{P2} and Assumption \ref{A2}, it can be derived that $\|M_0(q)\| \leq \overline \sigma_{m0}$,
and $\|h(t)\| \leq \overline \sigma_1 + \overline \sigma_2 \|q\| + \overline \sigma_3 \|\dot q\|^2$,
where $\overline \sigma_{m0}$, $\overline \sigma_1$, $\overline \sigma_2$, and $\overline \sigma_3$ are positive constants.
Design the PTSM vector as
\begin{equation}
s =  \dot e  + \frac{{{{\left[\kern-0.30em\left[ {{\textbf{1}_n} \!+\! {{\left[\kern-0.15em\left[ e
 \right]\kern-0.15em\right]}^2}}
 \right]\kern-0.30em\right]}^{\frac{3}{2}}}}}{{{{\mathcal T}_s}(1 - \gamma )}} \!\circ {\rm sig}{\left( {e \circ\! {{\left[\kern-0.30em\left[ {{\textbf{1}_n} \!+\! {{\left[\kern-0.15em\left[ e
 \right]\kern-0.15em\right]}^2}}
 \right]\kern-0.30em\right]}^{ - \frac{1}{2}}}} \right)^\gamma },
\end{equation}
where $\mathcal T_s > 0$, and $0 < \gamma < 1$.
For simplification, $M_0$, $C_0$, and $g_0$ are applied to respectively denote $M_0(q)$, $C_0(q,\dot q)$, and $g_0(q)$ in the following presentation.
Then the PTSM controller $\tau (t) = \tau_{eq} + \tau_s$ is presented as follows:
\begin{align}
&{\tau _{eq}} = -{M_0}\!\!\left[\! \frac{{3e \!\circ\! {{\left[\kern-0.30em\left[\! {{\textbf{1}_n} \!+\! {{\left[\kern-0.15em\left[ e
 \right]\kern-0.15em\right]}^2}}
 \!\right]\kern-0.30em\right]}^{\frac{1}{2}}} \!\!\circ\! \dot e}}{{{{\mathcal T}_s}(1 - \gamma )}} \!\circ\! {\rm sig}{{\left(\!\! {x \!\circ\! {{\left[\kern-0.30em\left[\! {{\textbf{1}_n} \!+\! {{\left[\kern-0.15em\left[ e
 \right]\kern-0.15em\right]}^2}}
 \!\right]\kern-0.30em\right]}^{ - \frac{1}{2}}}} \!\right)}^\gamma } \right.\nonumber\\
 &\qquad\;\left.+ \frac{{\gamma \dot e}}{{{{\mathcal T}_s}(1 \!-\! \gamma )}} \!\circ\! {{\left[\kern-0.35em\left[\! {\left| e \right| \!\circ\! {{\left[\kern-0.30em\left[ {{\textbf{1}_n} \!+\! {{\left[\kern-0.15em\left[ e
 \right]\kern-0.15em\right]}^2}}
 \right]\kern-0.30em\right]}^{ - \frac{1}{2}}}} \!\right]\kern-0.35em\right]}^{\gamma  \!-\! 1}} \right] \!+\! {C_0}\dot q \!+\! {g_0}, \label{e8}\\
&{\tau _s} =  - \frac{\pi }{{\rho {{\mathcal T}_c}}}\left( {\hat \sigma _{m0}^{1 - \frac{\rho }{2}} + \hat \sigma _{m0}^{1 + \frac{\rho }{2}}{{\left\| s \right\|}^{2\rho} }} \right)\frac {s}{{{\left\| s \right\|}^{ \rho }}} - C_0 s \nonumber\\
&\qquad- \left[ {\left( {{{\overline \sigma }_1} \!+\! {{\overline \sigma }_2}\left\| q \right\| \!+\! {{\overline \sigma }_3}{{\left\| {\dot q} \right\|}^2}} \right){I_{\rm{n}}} \!+\! {{\rm K}_d}} \right]{\mathop{\rm sgn}} (s) \label{e9},
\end{align}
where $\hat \sigma_{m0} = \frac{1}{2} \overline \sigma_{m0}$,
 $\mathcal T_c > 0$, $0 < \rho <1$, and ${\rm K}_d$ is a diagonal positive-definite gain matrix.
Then, substituting (\ref{e8}) and (\ref{e9}) into (\ref{e10}) yields the following closed-loop system,
\begin{align}
\dot s = & \!- M_0^{ - 1} \!\left\{\frac{\pi }{{\rho {{\mathcal T}_c}}}\left(\! {\hat \sigma _{m0}^{1 - \frac{\rho }{2}} \!\!+\! \hat \sigma _{m0}^{1 + \frac{\rho }{2}}{{\left\| s \right\|}^{2\rho} }} \!\right)\! \frac{s}{{{\left\| s \right\|}^\rho}} \!-\!d(t) \!-\! h(t)  \right.\nonumber\\
         &\!+\! C_0s \!+ \!\!\left.\left[ {\left( {{{\overline \sigma }_1} \!+\! {{\overline \sigma }_2}\!\left\| q \right\| \!+\! {{\overline \sigma }_3}\!{{\left\| {\dot q} \right\|}^2}} \right)\!{I_n} \!+\! {{\rm K}_d}} \right]{\mathop{\rm sgn}} (s)   \!\right\} \!-\! \alpha_r \label{e11}
\end{align}

\begin{theoremx}\label{T3}
Suppose that Assumptions \ref{A2} and \ref{A3} hold.
Utilizing the PTSM controller (\ref{e8}) and (\ref{e9}) for (\ref{e7}),
if
\begin{align}
&\mathcal T_c >0, \quad \mathcal T_s >0, \label{e12}\\
& \lambda_{\min}({{\rm K}_d}) \geq \overline \sigma_d + \overline \sigma_{m0}\overline \sigma_\alpha, \label{e13}
\end{align}
then the zero-error predefined-time stabilization problem for the robot manipulator can be addressed within $t \leq  \mathcal T_c + \mathcal T_s$.
\end{theoremx}

\vspace{0.5em}
\begin{proof}
Construct $V(s) = \frac{1}{2} s^T M_0 s$ as the Lyapunov function for (\ref{e11}).
According to Property \ref{P1}, $V(s)$ is positive-definite.
Then, differentiating $V(s)$ along (\ref{e11}) yields that
\begin{align}
\dot V(s) &= s^T M_0 \dot s + \frac {1}{2} s^T \dot M_0 s\nonumber \\
          &\leq \!\!- \frac{\pi }{{\rho {{\mathcal T}_c}}}\!\!\left(\! {\hat \sigma _{m0}^{1 - \frac{\rho }{2}}{{\left\| s \right\|}^{2 - \rho }} \!\!+\! \hat \sigma _{m0}^{1 + \frac{\rho }{2}}{{\left\| s \right\|}^{2 + \rho }}} \!\right) \!\!+\!\frac{1}{2}s^T(\dot M_0 \!-\! 2 C_0)s \nonumber\\
          &\!\quad- \left( {{{\overline \sigma }_1} + {{\overline \sigma }_2}\left\| q \right\| + {{\overline \sigma }_3}{{\left\| {\dot q} \right\|}^2} + {\lambda _{\min }}({{\rm{K}}_d})} \right)\left\| s \right\| \nonumber\\
          &\!\quad+ \left( {\left\| {h(t)} \right\| + \left\| {d(t)} \right\| +\left\| M_0\right\| \left\| {{\alpha _d}} \right\|} \right)\left\| s \right\| \nonumber \\
          &\!\leq - \frac{\pi }{{\rho {{\mathcal T}_c}}}\left( {\hat \sigma _{m0}^{1 - \frac{\rho }{2}}{{\left\| s \right\|}^{2 - \rho }} + \hat \sigma _{m0}^{1 + \frac{\rho }{2}}{{\left\| s \right\|}^{2 + \rho }}} \right)\nonumber\\
          &\!\quad- \left( {{\lambda _{\min }}({{\rm{K}}_d}) - \overline \sigma_d - \overline \sigma_{m0} \overline \sigma_\alpha} \right)\left\| s \right\| \nonumber \\
          &\!\leq - \frac{\pi }{{\rho {{\mathcal T}_c}}}\left( {\hat \sigma _{m0}^{1 - \frac{\rho }{2}}{{\left\| s \right\|}^{2 - \rho }} + \hat \sigma _{m0}^{1 + \frac{\rho }{2}}{{\left\| s \right\|}^{2 + \rho }}} \right),
\end{align}
where  $\rho \in (0,1)$ is predefined right after (\ref{e9}),
Properties \ref{P1}-\ref{P2} and (\ref{e13}) has been applied to obtain the above inequality.
Note that $V(s) \leq \hat \sigma_{m0} \|s\|^2$ due to that $\|M_0(q)\| \leq 2\hat \sigma_{m0}$.
It then follows that
\begin{equation}
\dot V(s) \leq - \frac{\pi }{{\rho {{\mathcal T}_c}}}\left( {{V^{1 - \frac{\rho }{2}}}(s) + {V^{1 + \frac{\rho }{2}}}(s)} \right).
\end{equation}
Thus, based on Lemma \ref{L1} and the analysis in Theorem \ref{T1},
it can be derived that the tracking errors $e$ and $\dot e$ converge to the origin within the predefined time $t \leq \mathcal T_f = \mathcal T_s + \mathcal T_c$.
This completes the proof.
\end{proof}

\vspace{0.5em}
\begin{remarkx}
Note that the tracking errors in \cite{Munoz-Vazquez} finally converge to a bounded set within the setting time,
due to the adoption of the linear sliding surface.
Thus, it is of great significance to design the PTSM surface (\ref{e1}),
based on which the PTSM controller is proposed to successfully solve the zero-error predefined-time tracking problem for robot manipulators.
\end{remarkx}

\vspace{0.5em}
\begin{remarkx}
Inspired by \cite{NingBD}, we further give the following TBG-based controller to address the practical predefined-time tracking problem for robot manipulators: $\tau = \tau_{eq} + \tau_s$,
where $\tau_{eq}$ is designed as same as (\ref{e8}) and $\tau_s$ is designed as
\begin{align}
{\tau _s} =   &- \left[ {\left( {{{\overline \sigma }_1} + {{\overline \sigma }_2}\left\| q \right\| + {{\overline \sigma }_3}{{\left\| {\dot q} \right\|}^2}} \right){I_{\rm{n}}} + {{\rm{K}}_d}} \right]{\mathop{\rm sgn}} (s) - {C_0}s \nonumber \\
              &- \frac{{{\hat \sigma }_{m0}}{\dot \varepsilon (t)}}{{1 - \varepsilon (t) + \epsilon}}s, \label{e14}
\end{align}
where $\varepsilon(t)$ is the TBG,
$\hat \sigma_{m0}$ is the same as that in (\ref{e9}),
$0< \epsilon \ll 1$,
and ${\rm K}_d$ is a diagonal positive-definite gain matrix.
\end{remarkx}

\begin{figure}[t]
\centering
\makebox{\includegraphics[width=7.5cm]{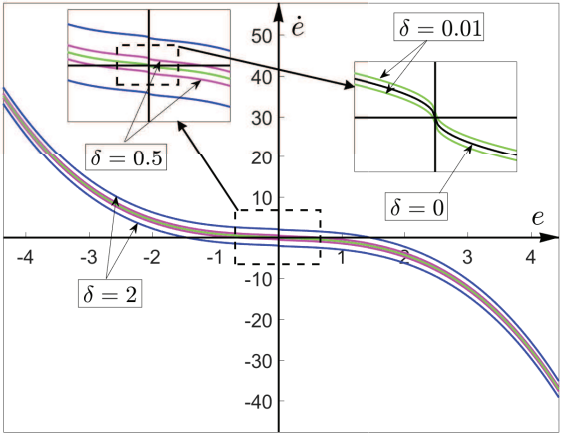}}
\caption{\label{F-9} The phase plot of the error system using the TBG-based controller.}
\end{figure}

\vspace{0.5em}
\begin{corollaryx}\label{C1}
Suppose that Assumptions \ref{A2} and \ref{A3} hold.
Utilizing the TBG-based controller (\ref{e8}) and (\ref{e14}) for (\ref{e7}),
if $\varepsilon(t)$ is give as in Lemma \ref{L2}, and
\begin{align}
&\mathcal T_c >0, \quad \mathcal T_s > 0, \nonumber\\
&\lambda_{\min}({\rm K }_d) \geq \overline \sigma_d + \overline \sigma_{m0} \overline \sigma_\alpha,
\end{align}
where $\mathcal T_c$ is the control parameter in $\varepsilon(t)$,
then the practical predefined-time stabilization problem of the robot manipulator can be addressed within $t \leq \mathcal T_c+ \mathcal T_s$.
\end{corollaryx}

\vspace{0.5em}
\begin{proof}
Substituting (\ref{e8}) and (\ref{e14}) into (\ref{e7}),
and differentiating $s$ along (\ref{e7}) yield below closed-loop system:
\begin{align}
\dot s =  & - M_0^{ - 1}\left\{ \frac{ {{\hat \sigma }_{m0}} {\dot \varepsilon (t)}}{{1 - \varepsilon (t) + \epsilon }}s - d(t) - h(t) + {C_0}s \right. \nonumber \\
          & \left.+ \left[ {\left( {{{\overline \sigma }_1} + {{\overline \sigma }_2}\left\| q \right\| + {{\overline \sigma }_3}{{\left\| {\dot q} \right\|}^2}} \right){I_{\rm{n}}} + {{\rm{K}}_d}} \right]{\rm{sgn}}(s) \right\} - {\alpha _r}.
\end{align}
Select $V(s) = \frac{1}{2} s^T M_0 s$.
It then follows that
\begin{align}
\dot V(s)  & = {s^T}{M_0}\dot s + \frac{1}{2}{s^T}{{\dot M}_0}s \nonumber \\
           & \le  - \frac{{{{\hat \sigma }_{m0}} \dot \varepsilon (t)}}{{1 \!-\! \varepsilon (t) \!+\! \epsilon}}{\left\| s \right\|^2} \!-\! \left( {{\lambda _{\min }}({{\rm{K}}_d}) \!-\! {{\overline \sigma }_d} \!-\! {{\overline \sigma }_{m0}}{{\overline \sigma }_\alpha }} \right)\left\| s \right\| \nonumber \\
           & \leq  - \frac{{\dot \varepsilon (t)}}{{1 - \varepsilon (t) + \epsilon}}V(s).
\end{align}
Based on Lemma \ref{L2},
it can be derived that $V(s)$ converges to a bounded set $\Omega = \{V(s)| V(s) \leq \frac{\epsilon}{1+\epsilon} V(s(0))\}$ at $\mathcal T_c$.
It then follows from $V(s) \geq \frac{\lambda_{\min}(M_0)}{2}\|s\|^2$ that ${\lim _{t \to {{\mathcal T}_c}}}\left\| s \right\| \le \delta$, where $\delta = \sqrt {\frac{2\epsilon}{{{\lambda _{\min }}({M_0})(1 + \epsilon)}}V(s(0))} $.
As is shown in Fig.\ref{F-9},
the error vectors will converge to a bounded neighborhood of the origin, which is related with $\delta$, in a predefined time $\mathcal T_s$.
To be specific,
the less $\delta$, the higher tracking performance we derive.
Besides, it is noteworthy that $\delta$ can be adjusted by selecting proper $\epsilon$.
Then, we can choose a small enough parameter $\epsilon$ to derive satisfactory tracking performance.
Thus, it can be concluded that the practical predefined-time stabilization problem of robot manipulators is addressed with the predefined time being $\mathcal T_c + \mathcal T_s$.
This completes the proof.
\end{proof}

\vspace{0.5em}
\begin{remarkx}
Redesign the nonlinear hitting control law $\tau_s$ as
\begin{align}
\tau_s =&  - {C_0}s - \left[ {\left( {{{\overline \sigma }_1} + {{\overline \sigma }_2}\left\| q \right\| + {{\overline \sigma }_3}{{\left\| {\dot q} \right\|}^2}} \right){I_{\rm{n}}} + {{\rm K}_d}} \right]{\mathop{\rm sgn}} (s) \nonumber\\
        &-  \alpha {\hat \sigma_{m0}^{\frac{m_1+n_1}{2n_1}}} {s^{\frac{{{m_1}}}{{{n_1}}}}} - \beta {\hat \sigma_{m0}^{\frac{m_2+n_2}{2n_2}}} {s^{\frac{{{m_2}}}{{{n_2}}}}}, \label{e15}
\end{align}
where $\hat \sigma_{m0}$ is the same as that in (\ref{e9}), $\alpha, \beta > 0$, $m_1, n_1, m_2$ and $n_2$ are positive integers with $m_1 > n_1$ and $m_2 < n_2$.
The main results thus can degrade into fixed-time convergence,
and the corresponding  corollary is given as follows.
\end{remarkx}

\vspace{0.5em}
\begin{corollaryx}
Suppose that Assumptions \ref{A2} and \ref{A3} hold.
Utilizing the TSM controller (\ref{e8}) and (\ref{e15}) for (\ref{e7}),
if (\ref{e13}) hold and $\mathcal T_s > 0$,
then $e_i$ and $\dot e_i$ converge to zero in fixed-time,
and the settling time function is given as
\begin{align}
\mathcal T \leq \mathcal T_s + \frac{2n_1}{{\alpha \left( {{m_1}-{n_1} } \right)}} + \frac{{{n_2} + {m_2}}}{{\beta \left( {{n_2} - {m_2}} \right)}}.
\end{align}
\end{corollaryx}

\vspace{0.5em}
\begin{proof}
Substituting (\ref{e8}) and (\ref{e15}) into (\ref{e7}) yields the following closed-loop system,
\begin{align}
\dot s =  & -\! M_0^{ - 1}\!\left\{ \!\alpha {\hat \sigma_{m0}^{\frac{m_1+n_1}{2n_1}}} \!\!{s^{\frac{{{m_1}}}{{{n_1}}}}} \!\!+\! \beta {\hat \sigma_{m0}^{\frac{m_2+n_2}{2n_2}}} \!\!{s^{\frac{{{m_2}}}{{{n_2}}}}} \!\!-\! d(t) \!-\! h(t) \!+\! {C_0}s \right. \nonumber \\
          & \left.+\! \left[ {\left( {{{\overline \sigma }_1} + {{\overline \sigma }_2}\left\| q \right\| + {{\overline \sigma }_3}{{\left\| {\dot q} \right\|}^2}} \right){I_{\rm{n}}} + {{\rm{K}}_d}} \right]{\rm{sgn}}(s) \right\} - {\alpha _r}.\label{e16}
\end{align}
Construct the Lyapunov function as $V(s) = \frac{1}{2}s^T M_0 s$.
Then, differentiating $V(s)$ along (\ref{e16}) yields that
\begin{align}
\dot V(s)     \leq&  - \alpha \hat \sigma _{m0}^{\frac{{{m_1} + {n_1}}}{{2{n_1}}}}{\left\| s \right\|^{\frac{{{m_1} + {n_1}}}{{{n_1}}}}} - \beta \hat \sigma _{m0}^{\frac{{{m_2} + {n_2}}}{{2{n_2}}}}{\left\| s \right\|^{\frac{{{m_2} + {n_2}}}{{{n_2}}}}} \nonumber\\
              &- \left( {{\lambda _{\min }}({{\rm{K}}_d}) - {{\overline \sigma }_d} - {{\overline \sigma }_{m0}}{{\overline \sigma }_\alpha }} \right)\left\| s \right\| \nonumber \\
             \leq&  - \alpha {V^{\frac{{{m_1} + {n_1}}}{{2{n_1}}}}}(s) - \beta {V^{\frac{{{m_2} + {n_2}}}{{2{n_2}}}}}(s).
\end{align}
Thus, it can be concluded that $s$ converges to the origin within $t \leq \frac{2n_1}{{\alpha \left( {{m_1}-{n_1} } \right)}} + \frac{{{n_2} + {m_2}}}{{\beta \left( {{n_2} - {m_2}} \right)}}$.
It follows that $e$ and $\dot e$ converge to zero within $t \leq \mathcal T_s +  \frac{2n_1}{{\alpha \left( {{m_1}-{n_1} } \right)}} + \frac{{{n_2} + {m_2}}}{{\beta \left( {{n_2} - {m_2}} \right)}}$.
This completes the proof.
\end{proof}

\section{Simulation Examples}\label{sec5}
In this section, several simulation experiments are conducted to verify the effectiveness of the main results.
\subsection{Simulation for the uncertain second-order system}
We first conduct simulation experiment on uncertain second-order system to verify the predefined-time stability presented in Theorem \ref{T2}.

\vspace{0.5em}
\begin{example}
The designed PTSM controller (\ref{e4}) and (\ref{e5}) is used to stabilize the uncertain second-order system (\ref{e2}).
Choose the state vectors $\xi, \eta \in \mathbb R^2$,
and the control parameters as $\gamma = 0.5$, $\rho = 0.4$, $\mathcal T_s = 4$, $\mathcal T_c = 6$, and ${\rm K}_f = 10$.
Besides, the uncertain term $f(t,\xi (t),\eta (t))$ is selected as a random function bounded by $5$.
Ten groups of initial value $\xi(0)$ and $\eta(0)$ are randomly chosen in $[-15,15]$.
\end{example}

\begin{figure}[t]
\centering
\makebox{\includegraphics[width=8.5cm]{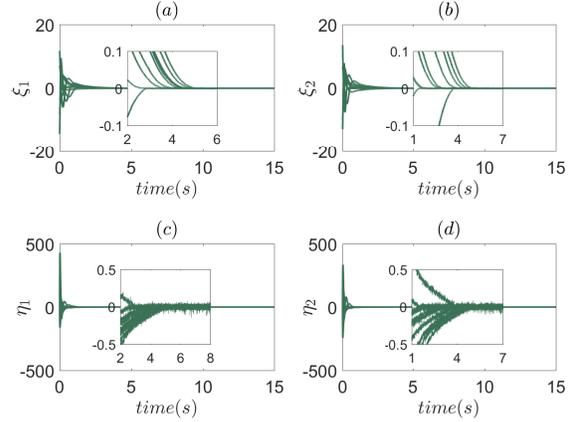}}
\caption{\label{F-1} Pictures (a) and (b) provides the evolution of $\xi$ for coordinates 1 and 2; pictures (c) and (d) provides the evolution of $\eta$ for coordinates 1 and 2.}
\end{figure}

\begin{figure}[t]
\centering
\makebox{\includegraphics[width=8.5cm]{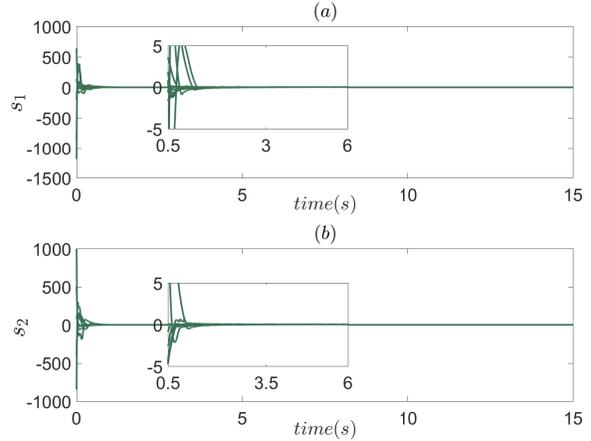}}
\caption{\label{F-2} Pictures (a) and (b) provides the evolution of the sliding mode vector $s$ for coordinates 1 and 2.}
\end{figure}

\textit{Simulation Results:} The simulation results are shown in Figs.\ref{F-1} and \ref{F-2}.
It can be observed in Fig \ref{F-1} that the states $\xi$ and $\eta $ converge to the origin within the predefined time $\mathcal T_f = \mathcal T_s + \mathcal T_c = 10s$ for any given initial values.
Similarly, Fig. \ref{F-2} shows the predefined-time convergence of the PTSM vector $s$.

\begin{figure}[t]
\centering
\makebox{\includegraphics[width=8.5cm]{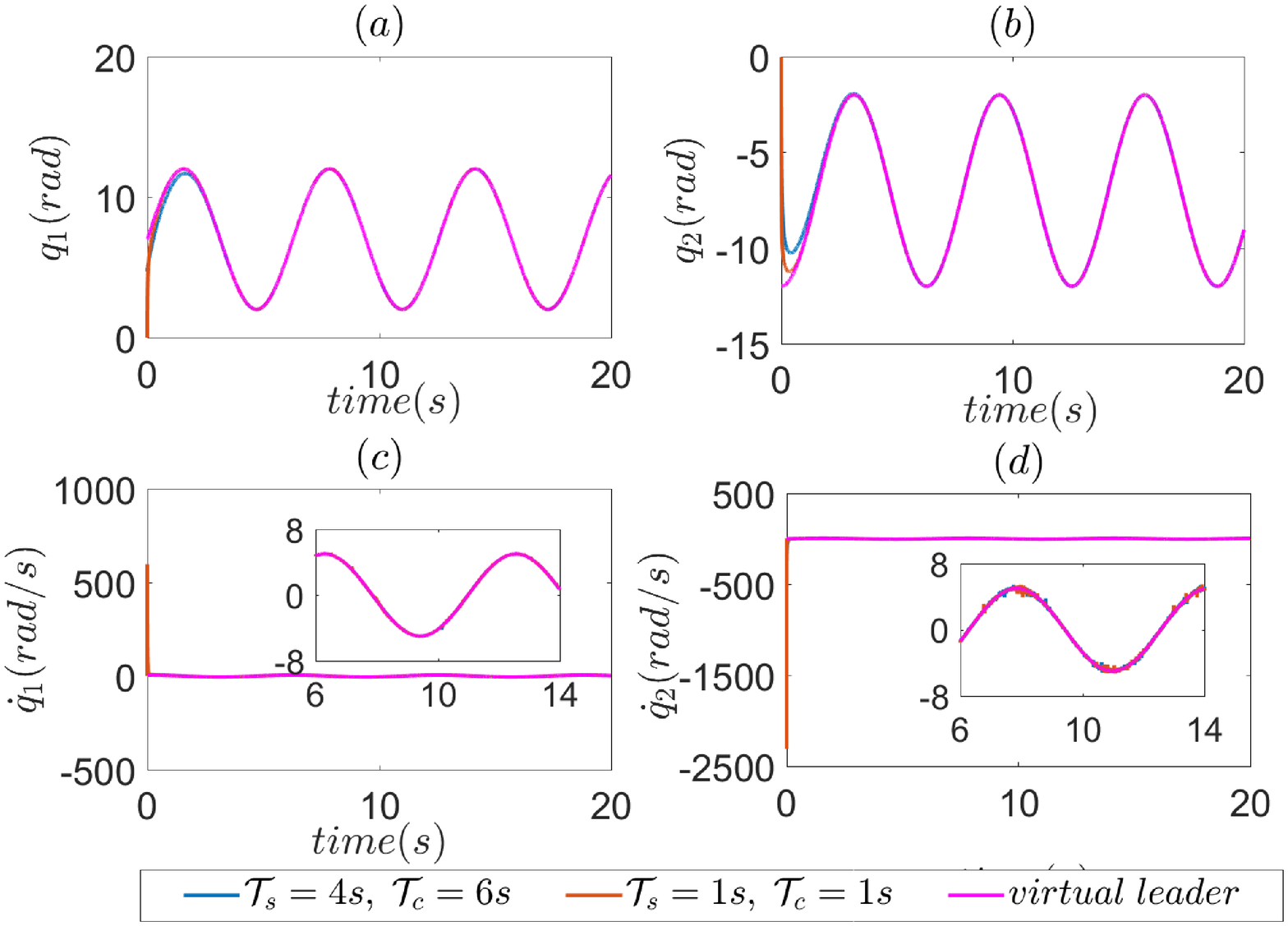}}
\caption{\label{F-3} Using the PTSM controller (\ref{e8}) and (\ref{e9}) for (\ref{e7}), pictures (a) and (b) provides the evolution of $q$ for coordinates 1 and 2; pictures (c) and (d) provides the evolution of $\dot q$ for coordinates 1 and 2.}
\end{figure}

\begin{figure}[t]
\centering
\makebox{\includegraphics[width=8.5cm]{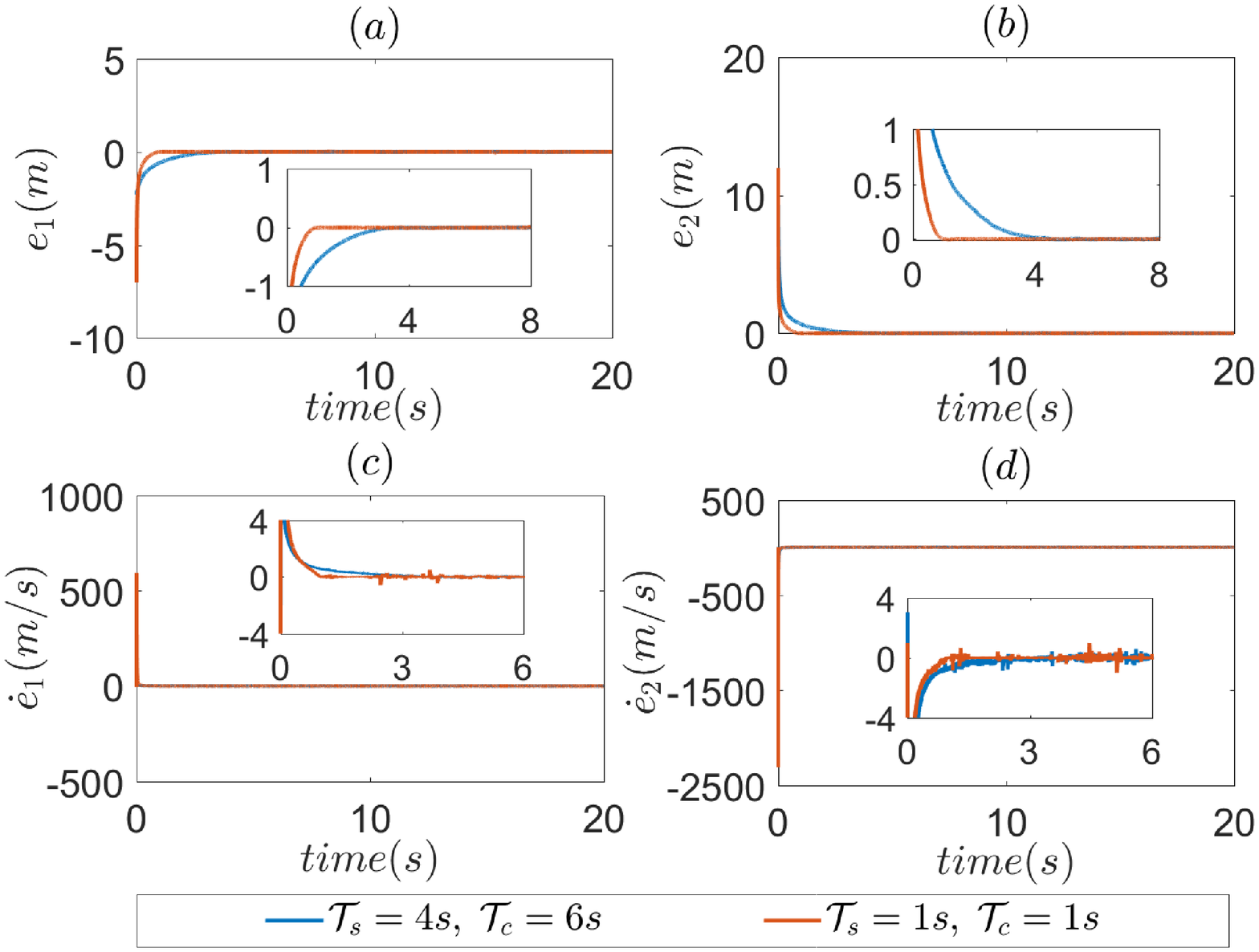}}
\caption{\label{F-4} Using the PTSM controller (\ref{e8}) and (\ref{e9}) for (\ref{e7}), pictures (a) and (b) provides the evolution of the tracking error $e$ for coordinates 1 and 2; pictures (c) and (d) provides the evolution of the tracking error $\dot e$ for coordinates 1 and 2.}
\end{figure}

\begin{figure}[t]
\centering
\makebox{\includegraphics[width=8.5cm]{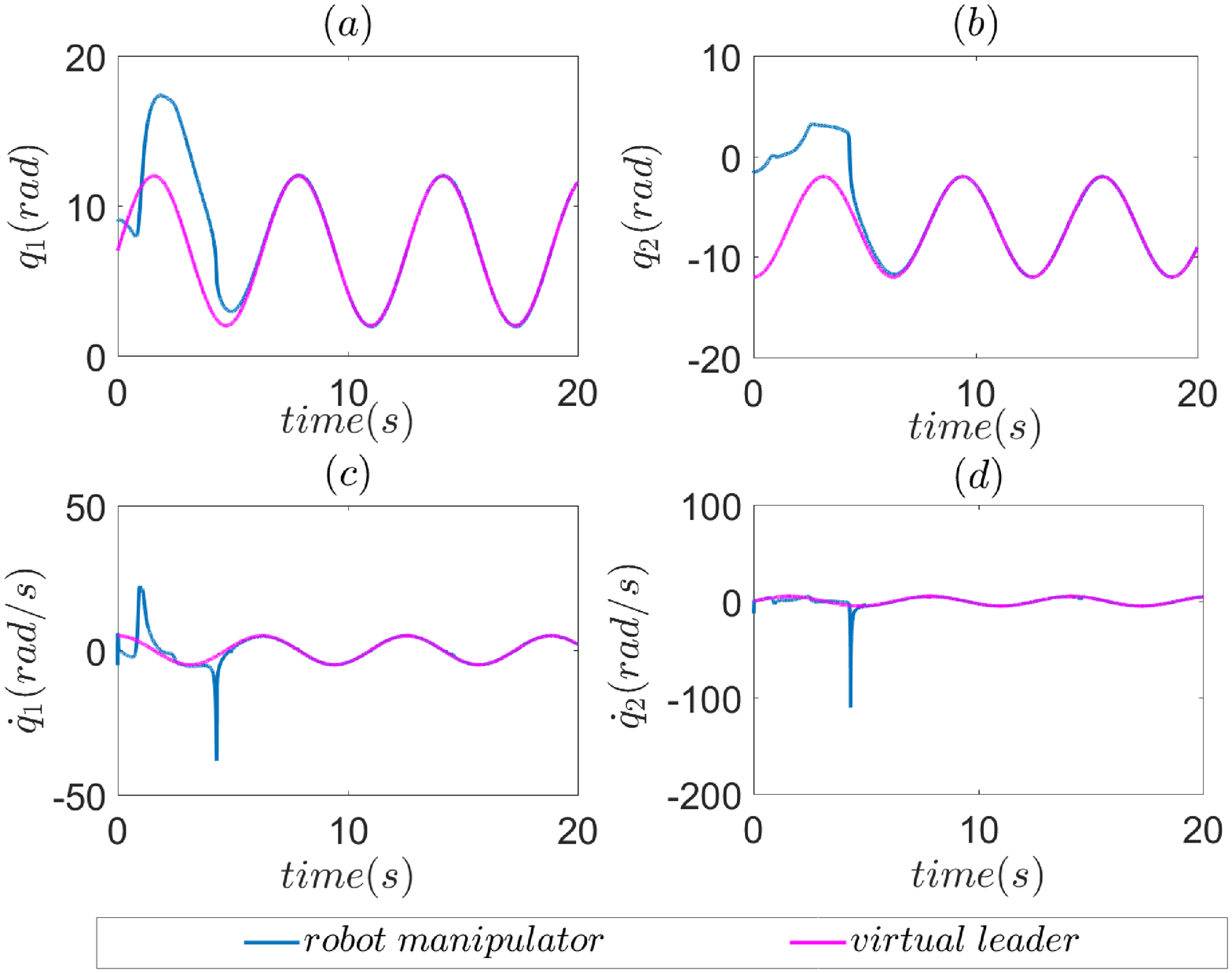}}
\caption{\label{F-5} Using the TBG-based controller (\ref{e8}) and (\ref{e14}) for (\ref{e7}), pictures (a) and (b) provides the evolution of $q$ for coordinates 1 and 2; pictures (c) and (d) provides the evolution of $\dot q$ for coordinates 1 and 2.}
\end{figure}

\begin{figure}[t]
\centering
\makebox{\includegraphics[width=8.5cm]{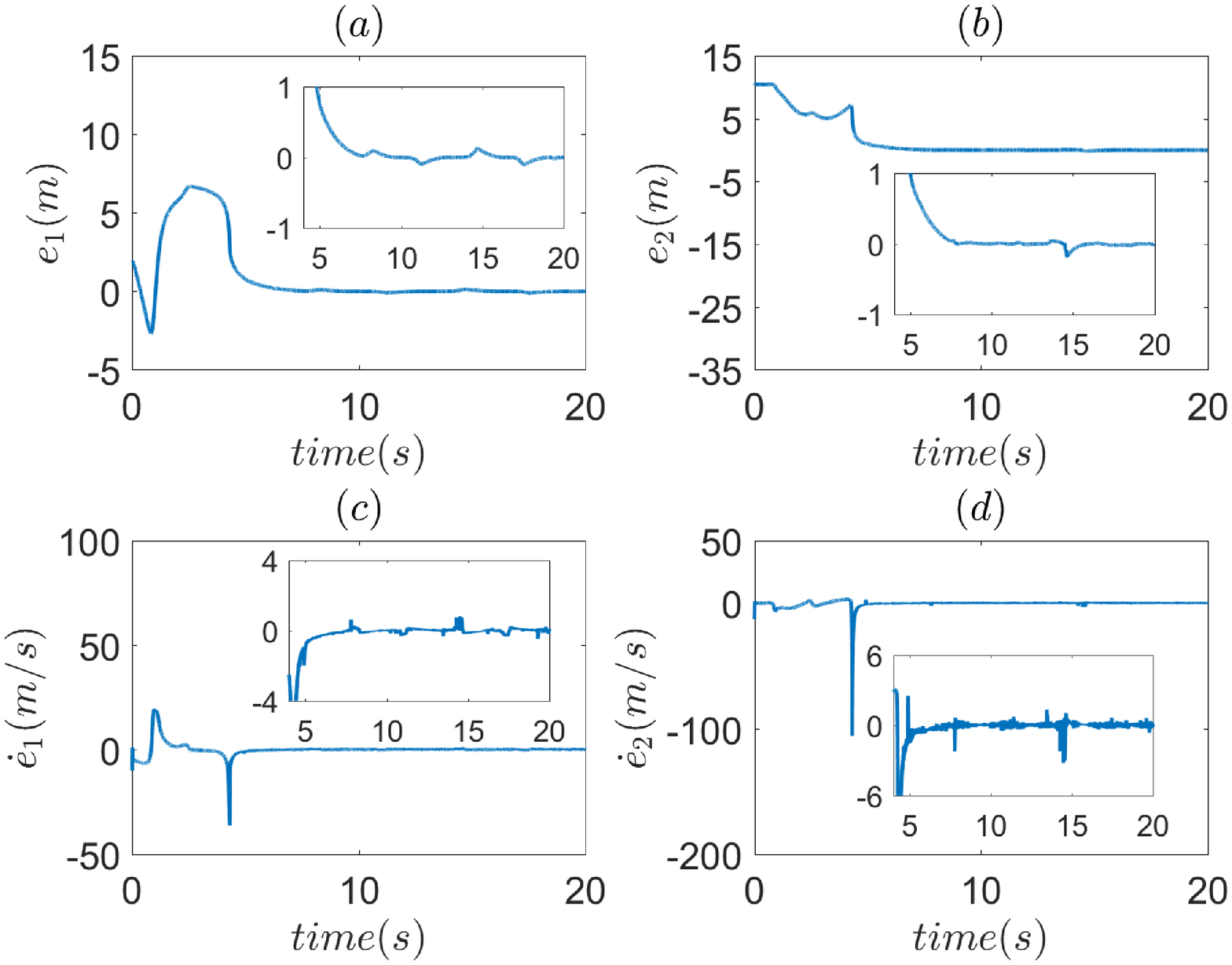}}
\caption{\label{F-6} Using the TBG-based controller  (\ref{e8}) and (\ref{e14}) for (\ref{e7}), pictures (a) and (b) provides the evolution of the tracking error $e$ for coordinates 1 and 2; pictures (c) and (d) provides the evolution of the tracking error $\dot e$ for coordinates 1 and 2.}
\end{figure}

\begin{figure}[t]
\centering
\makebox{\includegraphics[width=8.5cm]{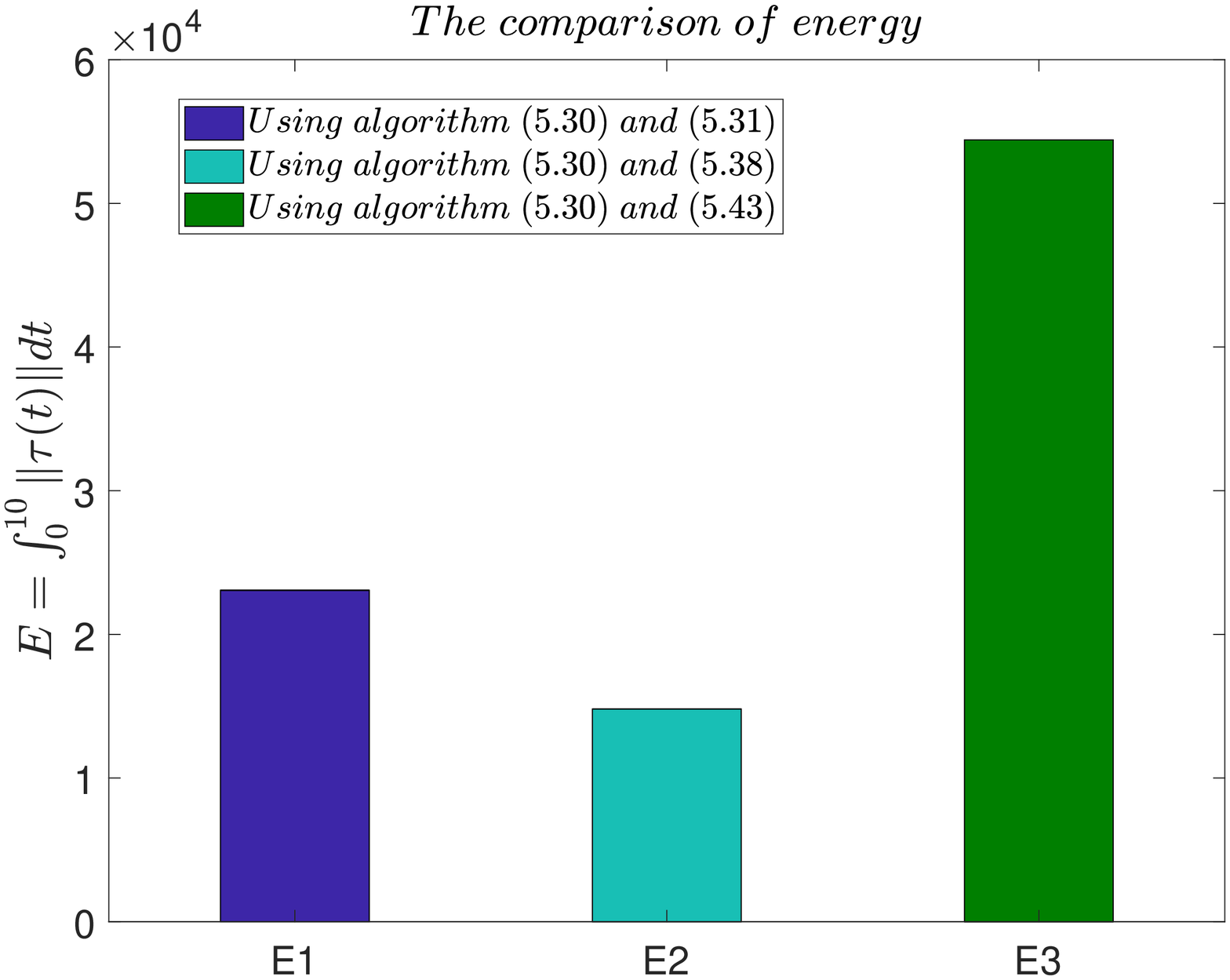}}
\caption{\label{F-7} The total energy consumed by using different control algorithms.}
\end{figure}

\subsection{Simulations for the robot manipulators}
In this subsection,
we select the 2-DOF robot manipulator to verify the effectiveness of the main results in Theorem \ref{T3} and Corollary \ref{C1}.
The specific dynamics is presented as follows:
\begin{align*}
&\left[\!\! {\begin{array}{*{20}{c}}
{M_{11}}&{M_{12}}\\
{M_{21}}&{M_{22}}
\end{array}} \!\!\right]\!\left[\!\! \begin{array}{l}
{{\ddot q}_{1}}\\
{{\ddot q}_{2}}
\end{array}\!\! \right] \!+\! \left[\!\! {\begin{array}{*{20}{c}}
{C_{11}}&{C_{12}}\\
{C_{21}}&{C_{22}}
\end{array}} \!\!\right]\!\left[\!\! \begin{array}{l}
{{\dot q}_{1}}\\
{{\dot q}_{2}}
\end{array} \!\!\right] \!+\! \left[\!\! \begin{array}{l}
{g_{1}}\\
{g_{2}}
\end{array}\!\! \right] \!=\! \left[\!\! \!\begin{array}{l}
{\tau _{1}}\\
{\tau _{2}}
\end{array} \!\!\right] \!+\! \left[\!\! \begin{array}{l}
{d_{1}}(t)\\
{d_{2}}(t)
\end{array} \!\!\right],
\end{align*}
where $M_{11} = p_{i1} + 2{p_{2}}\cos(q_{2})$,
$M_{12} = M_{21} = p_{3} + p_{2}\cos(q_{2})$,
$M_{22} = p_{3}$,
$C_{11} = -{p_{2}}\sin({q_{2}}){\dot q_{2}}$,
$C_{12} = -{p_{2}}\sin(q_{2})(\dot q_{1} + \dot q_{2})$,
$C_{21} = {p_{2}}\sin(q_{2}){\dot q_{1}}$,
$C_{22} = 0$,
$g_{1}= g{p_{4}}{\cos(q_{1})} + g{p_{5}}{\cos(q_{1}+q_{2})}$,
$g_{2} = g{p_{5}}{\cos(q_{1}+q_{2})}$,
$p_{1} = m_{1}{r_{1}^2} +  m_{2}(l_{1}^2 + r_{2}^2) +  I_{1} +  I_{2}$,
$p_{2} = m_{2} {l_{1}} { r_{2}}$,
$p_{3} = m_{2} {r_{2}^2} +  I_{2}$,
$p_{4} = m_{1} r_{1} +  m_{2}  l_{1}$,
$p_{5} = m_{2}  r_{2}$,
$ I_{1} = \frac{1}{3}{ m_{1}}  l_{1}^2$,
$I_{2} = \frac{1}{3}{ m_{2}} l_{2}^2$,
$g = 9.8m/s^2$ is the gravitational constant,
$m, l, r$  are the physical parameters, which are selected as
$m = [2.8, 1.8]^T {\rm kg} $,
$l = [3.8, 2.8]^T {\rm m} $,
and $r = \frac{1}{2}l$.
Besides the estimated value of the corresponding physical parameters are given as
$\hat m = [2.75, 1.85]^T {\rm kg} $,
$\hat l = [3.86, 2.74]^T {\rm m} $,
and $\hat r = \frac{1}{2}\hat l$.
On the other hand,
the trajectory of the referenced trajectory is chosen as
\begin{equation}
\left\{ \begin{array}{l}
{q_r} = {[7 + 5\sin (t), - 7 - 5\cos (t)]^T},\\
{{\dot q}_r} = {[5\cos (t),5\sin (t)]^T}.
\end{array} \right.
\end{equation}

\begin{example}\label{example2}
The PTSM controller (\ref{e8}) and (\ref{e9}) is used for system (\ref{e7}) to solve the zero-error predefined-time tracking problem.
The control parameters are selected as $\gamma = 0.5$, $\rho = 0.5$, $\overline \sigma_1 = 14$, $\overline \sigma_2 = 12$, $\overline \sigma_3 = 10$, ${\rm K}_d = 25 I_2$,
and $\overline \sigma_{m0} = 5$ to make the corresponding assumptions and sufficient conditions hold.
Besides, the external disturbance is selected as a random function bounded by $5$.
Besides, we select $\mathcal T_s = 4$, $\mathcal T_c = 6$, and $\mathcal T_s = 1$, $\mathcal T_c = 1$  to respectively carry out the simulation experiment with the initial values chosen randomly in $[-5,5]$.
\end{example}

\vspace{0.5em}
\begin{example}
The TBG-based controller (\ref{e8}) and (\ref{e14}) is used for system (\ref{e7}) to solve the practical predefined-time tracking problem.
The TBG $\varepsilon(t)$ is selected as follows to make the conditions in Lemma \ref{L2} hold,
\begin{equation*}
\varepsilon (t) = \left\{ \begin{array}{l}
\frac{{10}}{{{6^6}}}{t^6} - \frac{{24}}{{{6^6}}}{t^5} + \frac{{15}}{{{6^4}}}{t^4},\;\;t \in \left[ {0,{{\mathcal T}_c}} \right],\\
1,\qquad \qquad \qquad \qquad t \in \left( {{{\mathcal T}_c}, + \infty } \right),
\end{array} \right.
\end{equation*}
where $\mathcal T_c = 6$.
Besides, select $\epsilon = 0.1$, $\mathcal T_s = 4$,
and the remain settings are given as the same as that in Example \ref{example2}.
\end{example}

\textit{Simulation Results:} The simulation results are shown in Figs.\ref{F-3}-\ref{F-7}.
To be specific, Figs.\ref{F-3}-\ref{F-4} and Figs.\ref{F-5}-\ref{F-6} respectively show that the predefined-time tracking control for the 2-DOF manipulator can be realized by employing both the PTSM controller (\ref{e8}), (\ref{e9}) and the TBG-based controller (\ref{e8}), (\ref{e14}).
Besides, it can be seen from the tracking errors shown in Figs.\ref{F-4} and \ref{F-6} that using the PTSM controller (\ref{e8}), (\ref{e9}) derives higher tracking performance than using the TBG-based controller (\ref{e8}), (\ref{e14}),
which is consistent with the main results that using (\ref{e8}), (\ref{e9}) and (\ref{e8}), (\ref{e14}) respectively solve the zero-error and practical predefined-time tracking problems for the robot manipulator.
Next, comparisons of the total energy consumed by using the three different control algorithms in this paper have been made based on a metric of $E = \int_0^{10} {\left\| {\tau (t)} \right\|} dt$,
and the results is shown in Fig.\ref{F-7},
from which we can derive that the TBG-based controller can effectively reduce the energy consumption.
In conclusion, the main results presented in Theorems \ref{T1}-\ref{T3} are theoretically correct.

\section{Conclusion} \label{sec6}
This paper has presented a novel PTSM surface to design a new control scheme for stabilizing the tracking errors of the robot manipulator to the origin in a predefined time.
The sufficient conditions on the control parameters for guaranteeing zero-error predefined-time stability of the closed-loop system have been derived by carrying out the formal systemic analysis.
It is observed from the comparison studies that the PTSM control scheme provides better tracking performance while the TBG-based ones consumes less energy.
Finally, the simulation results have shown the satisfactory steady-error performance.
Future work will be focused on the predefined-time stabilization problem of robot manipulators in task space.

\end{document}